\documentclass[12pt]{aipproc}

\layoutstyle{8s}

\newcommand\doingARLO[2][]{%
  \ifx\mmref\undefined #1\else #2\fi
}

\newcommand{\la}{\stackrel{<}{ _{\sim}}}
\newcommand{\ga}{\stackrel{>}{ _{\sim}}}
\newcommand{\dddot}[1]{\stackrel{\ldots}{#1}}
\def\Ibar{\mathcal I}

\begin{document}

\title[Black Holes]{Black Holes: from Speculations to Observations}

\author{Thomas W. Baumgarte}{
  address={Department of Physics and Astronomy, Bowdoin College, 
    Brunswick, ME 04011},
  altaddress={Department of Physics, University of Illinois at 
    Urbana-Champaign, Urbana, IL 61801}}

\begin{abstract}
This paper provides a brief review of the history of our understanding
and knowledge of black holes.  Starting with early speculations on
``dark stars'' I discuss the Schwarzschild "black hole" solution to
Einstein's field equations and the development of its interpretation
from "physically meaningless" to describing the perhaps most exotic
and yet "most perfect" macroscopic object in the universe.  I describe
different astrophysical black hole populations and discuss some of
their observational evidence.  Finally I close by speculating about
future observations of black holes with the new generation of
gravitational wave detectors.
\end{abstract}


\maketitle

\section{Theoretical Considerations}

This paper summarizes a talk presented at the Albert Einstein Century
International Conference, held in Paris, France, in July 2005 to mark
the centennial of Einstein's "miracle" year 1905.  Strictly speaking,
black holes are a consequence of another Einstein miracle that did not
occur in 1905 and still had to wait ten years, namely general
relativity.  With his theory of special relativity, however, Einstein
had laid the ground work already in 1905 for his theory of general 
relativity.  And even more strictly speaking, speculations on black holes
predate even the special theory of relativity by over a century.

\subsection{Early Speculations}

In the late 1700's John Mitchell \cite{Mit1796} in England and Jean Simon
Laplace \cite{Lap1795} in France independently realized that celestial
bodies that are both small and massive may become invisible\footnote{An 
excellent review of the history of our understanding of compact objects can 
be found in \cite{Isr73}.}.   The
basis for this speculation is the observation that the escape
speed
\begin{equation}
v_{\rm esc} = \sqrt{ \frac{2 G M}{R}},
\end{equation}
where $M$ and $R$ are the stellar mass and radius, is independent of
the mass of the test particle.  Within Newton's particle theory of
light it seems quite reasonable that this should also apply to light,
in which case light can no longer escape the star if the escape speed
exceeds the speed of light,
\begin{equation}
v_{\rm esc} > c.
\end{equation}
Evidently this happens when
\begin{equation} \label{laplace}
G M > \frac{c^2 R}{2},
\end{equation}
meaning that stars with a large enough mass and a small enough radius
become ``dark''.  Laplace went on to speculate that such objects may
not only exist, but even in as great a number as the visible stars.
With the demise of the particle theory of light, however, these
speculations also lost popularity, and dark stars remained obscure
until well after the development of general relativity.

\subsection{General Relativity and the Schwarzschild solution}

In 1915 Albert Einstein published his field equations of general
relativity~\cite{Ein15},
\begin{equation} \label{field_gr}
G_{ab} = 8 \pi G T_{ab}.
\end{equation}
One could argue that superficially this equation may not be all that
different from the Newtonian field equation
\begin{equation} \label{field_newt}
\nabla^2 \phi = 4 \pi G \rho.
\end{equation}
The left-hand-side of the Newtonian field equation (\ref{field_newt})
features a second derivative of the Newtonian potential $\phi$, and
the right-hand-side contains matter densities $\rho$.  Quite similarly
the Einstein tensor $G_{ab}$ on the left-hand-side of Einstein's field
equations (\ref{field_gr}) contains second derivatives of the
fundamental object of general relativity, the spacetime metric
$g_{ab}$.  To complete the analogy, the stress-energy tensor $T_{ab}$
on the right-hand-side contains matter sources.  For the vacuum
solutions in which we are interested in this article, the stress-energy
tensor vanishes, $T_{ab} = 0$.  

Unfortunately the Einstein tensor $G_{ab}$ contains many lower-order,
non-linear terms, making Einstein's equations a complicated set of ten
coupled, quasi-linear equations for the ten independent components of
the spacetime metric $g_{ab}$.  Clearly, it is very difficult to find
meaningful exact solutions. Karl Schwarzschild, returning fatally
wounded from the battle fields of World War I, was nevertheless able
to derive a fully non-linear solution in spherical symmetry within a
year of Einstein's original publication \cite{Sch16}.  Written as a
line element $ds^2$, the spacetime metric $g_{ab}$ describing this 
solution is 
\begin{equation} \label{schwarzschild}
ds^2 = g_{ab} dx^a dx^b = 
- \left(1 - \frac{G}{c^2} \, \frac{2 M}{r} \right) dt^2
+ \left(1 - \frac{G}{c^2} \, \frac{2 M}{r} \right)^{-1} dr^2 
+ r^2 d\theta^2 + r^2 \sin^2
\theta d \phi^2.
\end{equation}
This solution is the direct analog of the Newtonian point-mass solution
\begin{equation}
\phi = \frac{G M}{r}
\end{equation}
and describes the strength of the gravitational fields, expressed by
the spacetime metric $g_{ab}$, created by a point mass $M$ at a
distance $r$.  The relativistic Schwarzschild solution
(\ref{schwarzschild}) is significantly more mysterious than its
Newtonian analog, though.  Particularly puzzling is the
``Schwarzschild radius''
\begin{equation} \label{r_ss}
r_{\rm S} = \frac{2 G M}{c^2} 
\end{equation}
at which the metric (\ref{schwarzschild}) becomes singular.  The
existence of this singularity obscured the physical interpretation of
this solution, and in fact Schwarzschild himself died believing that
his metric was physically irrelevant.

Other aspects also contributed to the fact that the astrophysical
significance of the Schwarzschild solution (\ref{schwarzschild})
remained unappreciated for decades.  For one thing it was not clear
how such an object could possibly form, even though in 1939
Oppenheimer and Snyder \cite{OppS39} published a remarkable analytic
calculation describing the ``continued gravitational contraction'' of
a dust ball that leaves behind the Schwarzschild metric
(\ref{schwarzschild}).  This calculation, serving as a crude model of
stellar collapse, made several simplifying assumptions: that the
matter has zero pressure (so that it can be described as dust), that
the angular momentum is zero, and that the spacetime is spherically
symmetric.  Critics maintained that in any realistic astrophysical
situation none of these assumptions would hold, and that any deviation
from these idealizations could easily halt the collapse, preventing
continued gravitational contraction.  The absence of angular momentum
seems particularly troubling.  Since the Schwarzschild solution does
not carry any angular momentum it was completely unclear how any
astrophysical object, which would necessarily carry {\em some} angular
momentum, could collapse and form such a solution.

Finally, no astronomical observations had revealed phenomena requiring
a gravitationally collapsed object as an explanation.  Given that at
least some of the astronomical community had been quite reluctant to
accept the much less exotic White Dwarfs as an astrophysical reality,
it is not surprising that a solution to Einstein's field equations
that became singular at finite radius was not immediately embraced as
a celestial object.

All three of these factors -- a lack of understanding of the
singularity at the Schwarzschild radius, general skepticism concerning
gravitational collapse, and the absence of astronomical observations
of gravitationally collapsed objects -- resulted in the fact that the
great astrophysical significance of Schwarzschild's solution
(\ref{schwarzschild}) remained unappreciated for almost 50 years.  All
of this changed in the 1960's, the ``golden age of black hole
physics''.

\section{The golden age of black hole physics}

The golden age of black hole physics was ushered in when advances in
our theoretical understanding of the Schwarzschild geometry and
gravitational collapse in general coincided with new astronomical
observations of highly energetic objects that clearly pointed to a
gravitationally collapsed object as their central engine.

\subsection{The Schwarzschild geometry}

To begin with, it became clear that the apparent singularity at the
Schwarzschild radius $r_{\rm S}$ (\ref{r_ss}) is simply a harmless
coordinate singularity, not completely unlike the poles of a sphere
when described in terms of longitude and latitude.  This was
demonstrated by \citet{Kru60}, who introduced a new coordinate
system\footnote{Interesting, at least one other regular coordinate
system had already been found by \citet{Pai21} and \citet{Gul22}.
\citet{Lem33} similarly concluded that the geometry at the
Schwarzschild radius must be regular, but apparently these conclusions
remained unnoticed.} that remains perfectly regular at $r_{\rm S}$.
Instead of being a mysterious singularity the Schwarzschild radius now
emerged as a point of no return: a one-way membrane called ``event
horizon'' through which nothing, not even light can leave the
collapsed region inside.  Inside this event horizon lurks a true
singularity at which the curvature of spacetime becomes
infinite\footnote{In a classical description of general relativity,
that is.  A self-consistent quantum theory of gravity will presumably
resolve this singularity.}.

\begin{figure}[t]
\includegraphics[width=4in]{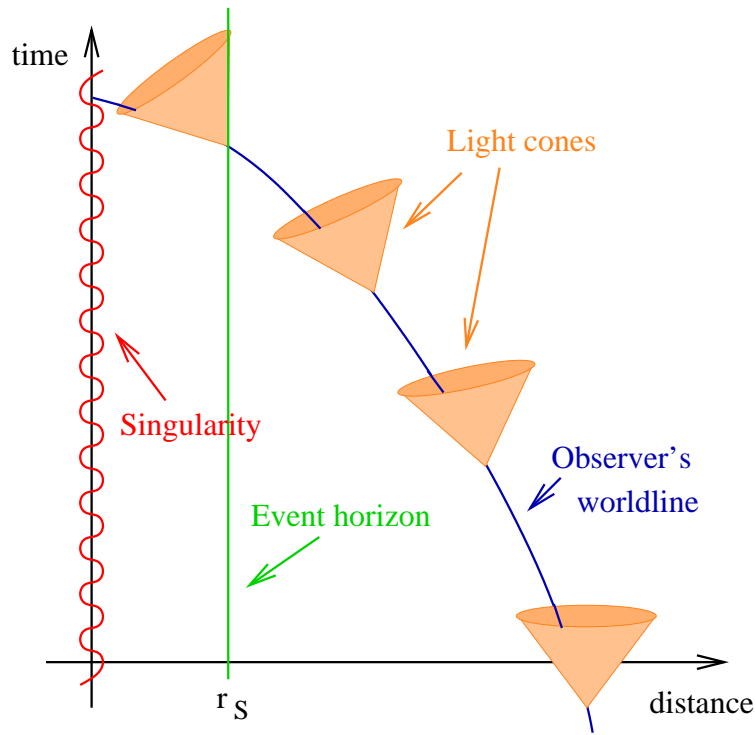}
\caption{Cartoon of a spacetime singularity surrounded by an event
horizon at $r_{\rm S} = 2M$.  The light cones of an observer falling
into the black hole become increasingly tilted, until they completely
tip over at the event horizon.}
\label{Fig1}
\end{figure}

Consider the unfortunate observer in the spacetime cartoon of Figure
\ref{Fig1}.  As long as he is far away from the event horizon an
``outgoing'' light ray emitted away from the horizon can propagate
toward larger distance almost as easily as an ``ingoing'' ray emitted
toward the horizon can propagate toward smaller distance.  At this
point the observer's local light cone, which is the section of
spacetime on which light propagates, is almost upright, and he can
easily send signals to a buddy even further away.  As our first
observer approaches the event horizon, however, his light cones become
increasingly tilted, and it becomes increasingly difficult for
outgoing light rays to actually move outward.  At the event horizon
the light cone tips over.  An ``outgoing'' light ray emitted exactly
on the event horizon will hover at the Schwarzschild radius forever,
and inside this point all light rays, even ``outgoing'' ones,
immediately move inward.  Nothing can emerge from the event horizon
and everything inside will soon reach the singularity that lurks at
the black hole's center.  Our observer will cross the event horizon
and reach the singularity in finite proper time.  As seen by his buddy
far away, his signals slowly fade away as he approaches the event
horizon.  Note also that the in-falling observer would not necessarily
observe anything special occurring at the event horizon.  Ultimately
the observer will be torn apart by the increasingly strong tidal
forces, but that may occur inside or outside the event horizon,
depending on the black hole's mass.  It is only through careful
experiments with flashlights, studying the properties of outgoing
light-rays, that the observer would be able to tell that he has
crossed an event horizon.

It is interesting to note that, by sheer coincidence, the location of
the event horizon $r_{\rm S}$ (\ref{r_ss}) as expressed in
Schwarzschild coordinates coincides exactly with the radius of a
``dark star'' (\ref{laplace}) as determined by Mitchell and Laplace.

\subsection{Gravitational Collapse}

Almost simultaneously with the improved understanding of the
Schwarzschild geometry, it became clear that gravitational collapse is
much more generic than was believed earlier.  In 1963 Roy Kerr
\cite{Ker63} discovered a generalization of the Schwarzschild solution
(\ref{schwarzschild}) that carries angular momentum.  With this 
discovery the argument that objects carrying angular momentum cannot
collapse gravitationally immediately collapsed itself.  The first 
numerical relativity simulations also demonstrated that pressure may
not be able to prevent gravitational collapse \cite{MayW66}.

Further progress in our understanding of gravitational collapse
arrived with a number of theorems.  Perhaps most importantly Roger
Penrose \cite{Pen65} showed that the formation of a spacetime
singularity is generic after a so-called trapped surface has formed.
Another set of theorems are collectively called the ``no-hair''
theorems\footnote{See, e.g., \citet{Isr67,Car71,Rob75}.}; in essence
these theorems state that black holes have no distinguishing features.
That is to say that Kerr's solution describing a rotating black hole
is the {\em only} solution describing a rotating black hole -- all
stationary black holes are Kerr black holes, parametrized only by
their mass $M$ and the angular momentum $J$\footnote{Strictly speaking
black holes can also carry a charge $Q$, but from an astrophysical
perspective this is irrelevant since any charge would very be
neutralized very quickly.}.  Black holes can be perturbed, but any
perturbation is quickly radiated away, leaving behind a Kerr black
hole.

This is a truly remarkable statement: it means that the structure of
black holes is {\em uniquely} determined by their mass and angular
momentum alone, completely independently of what formed the black hole
at first place.  This observation led Chandrasekhar to eloquently
conclude in his Nobel Lecture
\begin{quotation}
This is the only instance we have of an exact description of a
macroscopic object...  They are, thus, almost by definition, the most
perfect macroscopic objects there are in the universe.
\end{quotation}

\subsection{Observational Evidence}

The technical advances during World War II lead to the development of
radio astronomy in the post-war years.  Within a few years several
discrete sources had been detected, but with few exceptions the origin
of these sources remained a mystery.  It was generally believed that
the sources were otherwise dark ``radio stars'' in our galaxy, since
at extra-galactic distances they would have to be enormously
energetic.  This opinion started to change when the positioning of
these ``quasars'' improved, and some were identified with galaxies.  

The true breakthrough arrived when lunar occultations of the radio
source 3C273 lead to its identification with another object whose
optical spectrum showed a redshift of $z = 0.158$ -- clearly
establishing it as an extra-galactic object.  This realization is
documented in the remarkable March 16 issue of {\rm Nature}
\cite{HazMS63,Sch63,Oke63,GreM63}.  The same volume contains a
theoretical paper by \citet{HoyF63}, in which the authors
conclude\footnote{See also \citet{Gin61}.}
\begin{quotation}
Our present opinion is that only through the contraction of a mass of
$10^7 - 10^8 M_{\odot}$ to the relativity limit can the energies of
the strongest sources be obtained.
\end{quotation}
These events ushered in ``Relativistic Astrophysics'' as
a new field.  As a sign of the time the first {\em Texas Symposium on
Relativistic Astrophysics} was convened in December of 1963.  The new
field also attracted many new people into the field, including John
Wheeler.  In fact, it was John Wheeler who in 1967 coined the term
{\em black hole}, marking the transition from speculative ideas on
dark stars to the astrophysical reality of black holes\footnote{I
again refer to \citet{Isr73} for a much more complete account of the
developments described in this Section.}.

\section{Astrophysical Black Holes}

Clearly, we do not have any absolutely water-tight proof that black
holes exist in our universe.  However, we do have some extremely
convincing evidence that makes black holes by far the most
conservative explanation of the observed phenomena.

Observations clearly point to two different populations of black
holes.  One of these populations are ``stellar-mass black holes'',
which have masses in the order of 10 $M_{\odot}$; another group are
``supermassive black holes'', which have masses in the order of $10^6
- 10^9 M_{\odot}$.  I will discuss these two groups in more detail
below.  There is also some observational evidence for ``intermediate
mass black holes'' of about $10^3 - 10^4 M_{\odot}$, and finally there
are speculations on ``primordial black holes'' that would be left over
from the big bang.  Even though these different kinds of black holes
have vastly different masses, they are, from a mathematical
perspective, the exact same kind of animal: a Kerr solution to
Einstein's field equations with certain values for their mass and
angular momentum.  From an astrophysical perspective they differ not
only in their mass and angular momentum, but also in how we can
observe them and in how they form, i.e.~in their evolutionary history.

\subsection{Stellar-mass black holes: Cygnus X-1}

The prime example of a stellar-mass black hole is Cygnus
X-1\footnote{See \citet{Oda77} for a review and references.}, which
was first discovered in the data of the X-ray satellite Uhuru,
launched on Dec 12 1970.  The case for Cygnus X-1 as a black hole can
be summarized as follows\footnote{See Section 13.5 and Box 13.1 in
\citet{ShaT83} for a detailed discussion.}: To begin with, Cygnus X-1
shows very short time variations in the X-ray signal (in the order of
$\Delta t \sim 10$ ms and less).  This implies that Cyg X-1 is a very
small object, in the order of $R \la c \Delta t \la 10^6$ m.  It is
also the unseen binary companion to a 9th magnitude supergiant star
called HDE226868.  The Doppler curve of this star shows that the
binary has an orbital period of about 5.6 days; from the amplitude of
the Dopplershift we can determine the mass function $f$ to be about
$0.25 M_{\odot}$ \cite{GieB82}.  Combining this with the mass of
supergiant stars we can derive a lower limit on the mass of Cyg X-1,
\begin{equation}
M_{\rm Cyg~X-1} \ga 3.3 M_{\odot}.
\end{equation}
An independent argument involving some constraints on the binary's
orbit arrives at a similar limit.  

Given its size, Cyg X-1 has to be a ``compact object'': a white dwarf,
a neutron star, or a black hole.  Even under very conservative
assumptions both white dwarfs and neutron stars have maximum masses
safely below the lower limit of Cyg X-1's mass.  This leaves a black
hole as the most likely explanation.

Cyg X-1 is an example of a {\em stellar-mass} black hole, which form
the end-point of the evolutionary cycle for massive stars.  When
massive stars run out of nuclear fuel they can no longer support
themselves against gravitational contraction.  For stars with masses
larger than about 20 $M_{\odot}$ nothing can halt the subsequent
gravitational collapse, which therefore leads to prompt black hole
formation.  For stars with masses between about 8 and 20 $M_{\odot}$
the collapse can be halted when the density of the compressed stellar
material reaches nuclear densities.  The resulting shock-wave launches
a ``core-collapse'' supernova and leaves behind a newly formed neutron
star.  This neutron star may either remain stable, or it may form a
black hole at a later time.  A ``delayed collapse'' to a black hole
can be triggered by a variety of mechanisms, including fall-back of
matter and phase transitions in the neutron star interior.  Finally,
two neutron stars may collide.  If the remnant exceeds the maximum
allowed mass for rotating neutron stars this coalescence will also
lead to the formation of a stellar-mass black hole\footnote{Even if
the remnant exceeds the maximum allowed mass for {\em uniformly}
rotating neutron stars it may be stabilized temporarily by virtue of
{\em differential} rotation \cite{BauSS00}.  Dissipation of the
differential rotation, for example by magnetic coupling, then triggers
a delayed collapse}.

We currently know of about 20 confirmed black hole
binaries\footnote{See Table 4.1 in \citet{McCR06}.}, but presumably
that is only a tiny fraction of the total number of stellar-mass black
holes in our own galaxy.

\subsection{Supermassive black holes: Sagittarius A$^{*}$}

\begin{figure}[t]
\includegraphics[width=4in]{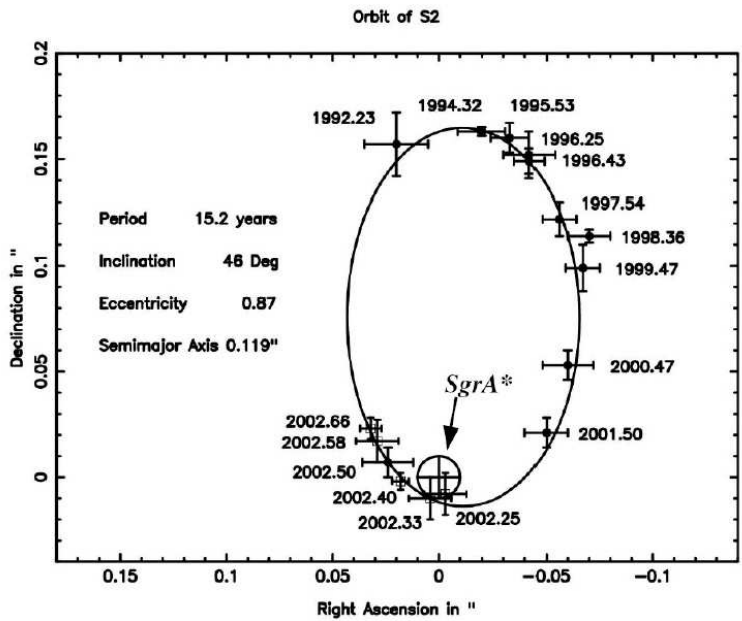}
\caption{Orbit of the star S2 around SgrA$^*$ (Figure from \cite{Schetal02}.).}
\label{Fig2}
\end{figure}

Perhaps the most convincing evidence for a black hole comes from the
center of our own galaxy, Sagittarius A$^{*}$
\cite{Schetal02,Gheetal03}.  Observations of our galactic center in
the near infrared reveal several stars that orbit a central object in
bound orbits\footnote{A beautiful movie animation of these orbits can
be found at {\tt www.mpe.mpg.de/ir/GC}.}.  Particularly compelling is
the orbit of the star S2, about two-thirds of which has now been
mapped with increasingly accurate positioning (see Figure \ref{Fig2}).
The emerging orbit is a Kepler orbit with a period $P$ of about $15$
years and a semi-major axis $a$ of about 4.62 mpc.  From Kepler's
third law we can conclude that the enclosed mass is
\begin{equation}
M = \frac{4 \pi^2 a^3}{G P^2} \approx 4 \times 10^6 M_{\odot}
\end{equation}
S2's orbit also has a significant eccentricity of $e = 0.87$.  At
pericenter its distance to the central object is only 124 AU.  This
implies that the central object harbors an enormous mass in a very
small volume.  The most conservative explanation for such an object --
which also must have remained stable over the lifetime of the galaxy
-- is again a black hole.

Sagittarius A$^{*}$ is an example of a {\em supermassive} black hole.
There is some very convincing evidence for supermassive black holes in
other galaxies as well -- for example the maser observations from NGC
4258 (also known as M106) \cite{Miyetal95}.  In fact, there is
evidence that supermassive black holes lurk at the cores of most
galaxies \cite{Ricetal98}.

It is less clear exactly how supermassive black holes form.  Many
different routes may lead to the formation of massive black holes in
active galactic nuclei\footnote{See, e.g., the flow chart in Fig.~1 of
\cite{Ree84}.}, but which of these routes nature tends to take is
still under debate.  A constraint comes from the recent observation of
quasars at redshift $z \approx 6$ in the Sloan Digital Sky Survey
\cite{Fanetal03}.  If these quasars are indeed powered by supermassive
black holes, this implies that the latter must have formed very
quickly in the early universe.  One model that may account for that is
accretion onto seed black holes that form in the collapse of
first-generation (Pop.~III) stars \cite{Sha05,VolR05}.

\subsection{Probing black hole properties}
\label{bhprop}

\begin{figure}[t]
\includegraphics[width=3in]{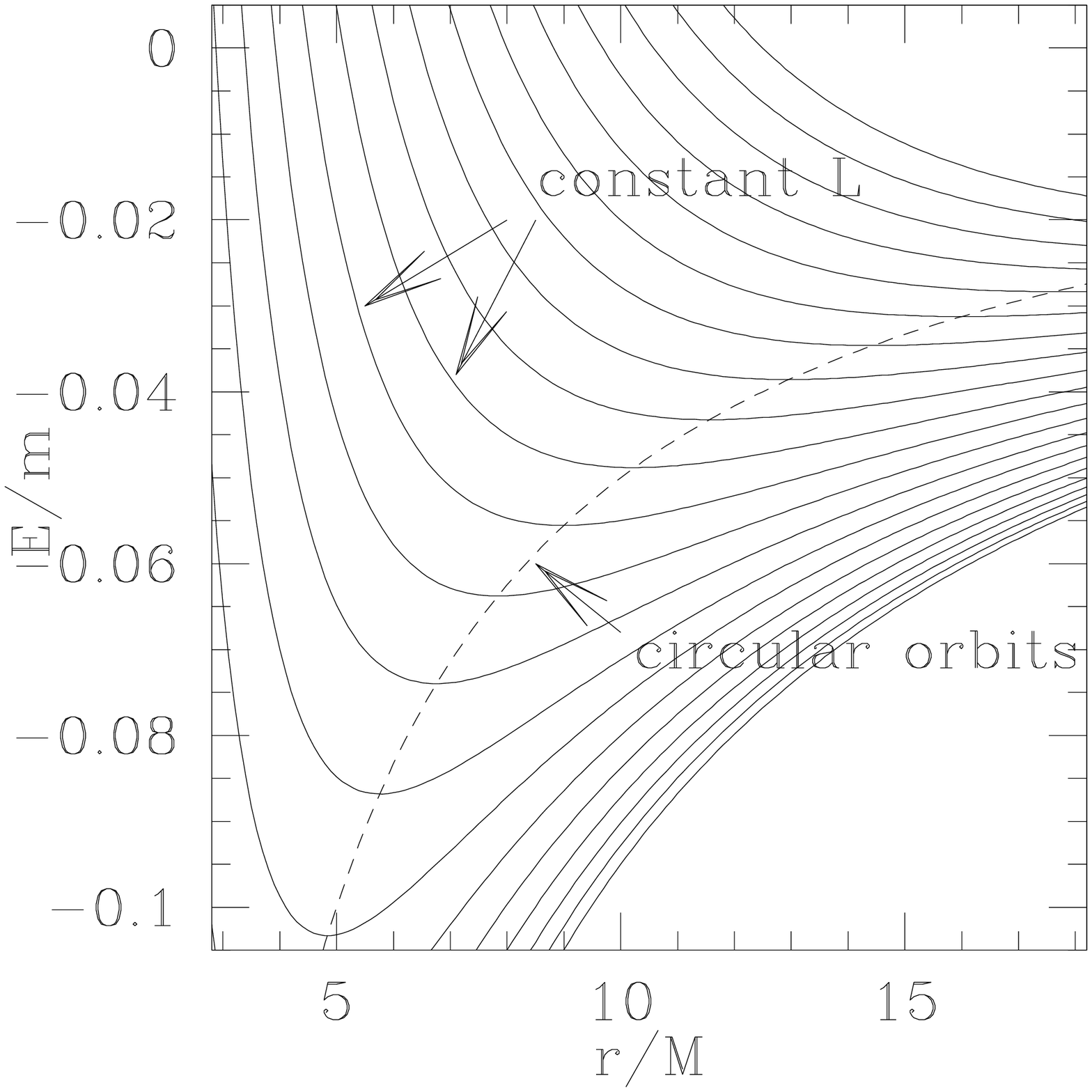}
\includegraphics[width=3in]{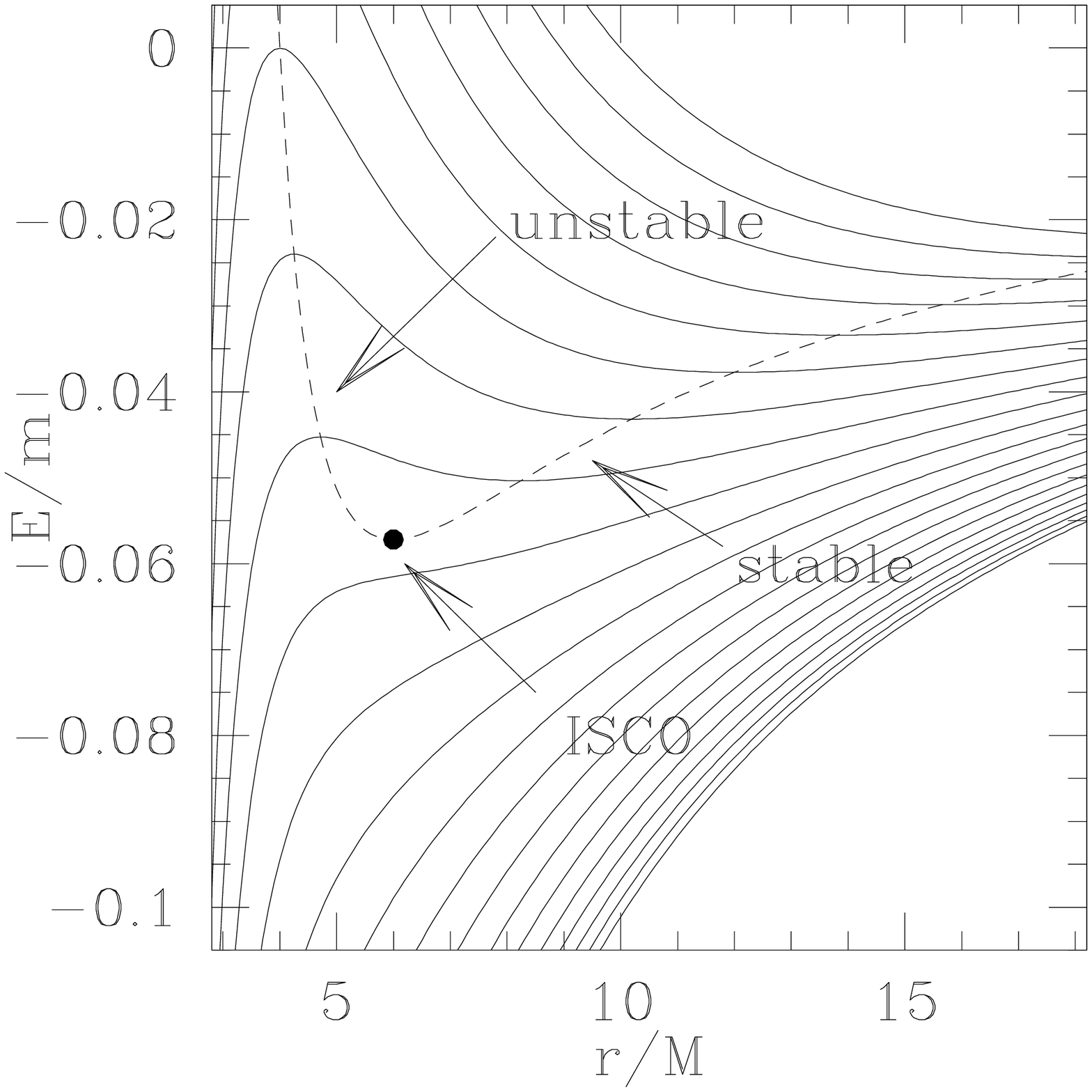}
\caption{The binding energy $E/mc^2$ as a function of separation $rc^2/GM$ of
a test mass $m$ in orbit about a Newtonian point mass $M$ (left panel)
and a Schwarzschild black hole of mass $M$ (right panel).  The solid
lines denote contours of constant orbital angular momentum $L$.
Extrema of these contours identify circular orbits, marked by the
dashed lines.  The circular orbits are stable if the extremum is a
minimum, otherwise they are unstable.}
\label{Fig3}
\end{figure}

Most observational evidence for black holes to date is based on the
argument ``a lot of mass in a tiny volume'', which leaves black holes
as the most conservative explanation.  Also, most observations to date
only provide information about the black hole mass, and not about the
angular momentum.  Clearly it would be desirable to go beyond that.

A number of efforts are under way to find observational evidence for
certain black hole characteristics.  One such characteristic is the
absence of a stellar surface.  X-rays emitted by black hole
candidates indeed shows some differences from that emitted by neutron
stars.  These differences can be explained in terms of an accretion
flow hitting the stellar surface in the case of neutron stars, but
freely falling through the horizon in case of the black
holes\footnote{See \citet{Nar05} for a review and references.}.
Another effort aims at resolving the event horizon of Sgr A$^*$, which
would show up as a ``black hole shadow'' since it absorbs all
radiation emitted behind it.

An interesting idea for the measurement of the black hole angular
momentum $J$ is based on the concept of the {\em Innermost Stable
Circular Orbit}, or ISCO for short.  Consider a test mass $m$ in orbit
about a point mass $M$.  A circular orbit can be found by identifying
an extremum of the test mass's binding energy $E$ at constant orbital
angular momentum $L$; a minimum corresponds to a stable circular
orbit, while a maximum corresponds to an unstable orbit.

In Newtonian physics the binding energy as a function of separation
$r$ is given by
\begin{equation} \label{Newt_energy}
\frac{E}{mc^2} = - \frac{G}{c^2} \, \frac{M}{r} + c^2 \frac{\tilde L^2}{2r^2},
\end{equation}
where $\tilde L = L/mc^2$.  To find an extremum of the binding energy
we differentiate with respect to $r$ at constant $\tilde L$ and set
the result to zero,
\begin{equation} \label{L}
\tilde L^2 = \frac{G M r}{c^4},
\end{equation}
which we recognize as Kepler's third law (since $\tilde L = \Omega
r^2/c^2$).  Evidently we can find a circular orbit for arbitrary $r$, and
the equilibrium energy of these circular orbits results from inserting
(\ref{L}) back into (\ref{Newt_energy}),
\begin{equation} \label{Newt_bind}
E_{\rm bind} = \frac{1}{2}\, \frac{G M m}{r}. 
\end{equation} 
We can also verify that all these extrema are minima, meaning that the
orbits are stable.  Alternatively, we can identify the orbits
graphically as in the left panel in Fig.~\ref{Fig3}.

For a test mass in orbit about a Schwarzschild black hole the binding
energy is given by
\begin{equation}
\frac{E}{mc^2} = \left( \left( 1 - \frac{G}{c^2} \,\frac{2 M}{r} \right)
\left( 1 + c^2 \frac{\tilde L}{r^2} \right) \right)^{1/2} - 1,
\end{equation}
where $r$ is the Schwarzschild radius.  Taking derivatives we would
find that circular orbits exist only for radii $r > 3 GM/c^2$.
Moreover, these orbits are stable only outside
\begin{equation}
r_{\rm ISCO} = \frac{6GM}{c^2}, 
\end{equation}
which therefore marks the ISCO for a test particle orbiting a
Schwarzschild black hole (see also the right panel in
Fig.~\ref{Fig3}).

The presence of an ISCO is of great relevance for black hole
observations because most of the emitted radiation is believed to
originate from an accretion disk.  In this accretion disk particles
follow almost circular orbits as they spiral toward the black hole.
Since circular orbits are unstable inside the ISCO, the accretion disk
can only exist outside $r_{\rm ISCO}$.  Accordingly, the Doppler shift
of spectral lines is limited by the speed of matter at the ISCO.  The
key point is that the location of the ISCO depends on the black hole's
angular momentum $J$; it is at $r_{\rm ISCO} = 6GM/c^2$ only for a
Schwarzschild black hole with $J =0$, and can get much closer to the
event horizon for a spinning Kerr black hole.  For a spinning black
hole the accretion disk may therefore extend closer to event horizon,
and the correspondingly higher speeds result in a greater broadening
of the emitted spectral lines.  Some results based on this idea have
been reported in \cite{Miletal04}.  

Even if successful, however, these techniques can only determine the
global parameters $M$ and $J$.  Clearly it would be desirable to map
out the local properties of the spacetime geometry around a black
hole.  Our best chance of doing that comes with gravitational wave
observations.

\section{Gravitational Radiation}

\subsection{Sources of gravitational radiation}

Maxwell's equations predict that accelerating charges emit
electromagnetic radiation.  For an electric dipole, the emitted power
is given by the Lamor formula
\begin{equation}
L_{\rm em} = \frac{2}{3 c^3} \ddot{d_i} \ddot{d_i},
\end{equation}
where $d_i$ is the electric dipole moment, $\ddot d_i$ its second time
derivative, and where we sum over repeated indices.

Einstein's equations similarly predict that accelerating masses emit
gravitational radiation.  One might expect that the power emitted from
a gravitational wave source is given by an expression similar to
Lamor's formula.  However, the analog of the electric dipole moment is
the mass dipole moment, and its first time derivative the total
momentum.  For an isolated system the total momentum is constant, so
that the second derivative of the dipole moment vanishes -- there is
no gravitational dipole radiation.  The first -- and often dominant
-- contribution to gravitational radiation comes from the quadrupole
term.  The equivalent of the Lamor formula in general relativity is
therefore
\begin{equation}
L_{\rm GW} = \frac{1}{5}\,\frac{G}{c^5}\,  
\langle \dddot{\Ibar}_{jk} \dddot{\Ibar}_{jk} \rangle
\end{equation}
Here $\Ibar_{ij}$ is the reduced quadrupole moment
\begin{equation} \label{J_quad_mom}
\Ibar_{jk} = \int \rho \left( x_j x_k - \frac{1}{3} \delta_{jk} r^2
\right) d^3 x
= \sum_A m_A \left( x_j x_k - \frac{1}{3} \delta_{jk} r^2
\right),
\end{equation}
the bracket $\langle \rangle$ denotes averaging over several
characteristic periods of the system, and the triple dot denotes the
third time derivative.  For a strong signal of gravitational radiation
we evidently need large and rapidly changing quadrupole moments, which
brings to mind binary systems.

\begin{figure}[t]
\includegraphics[width=6.5in]{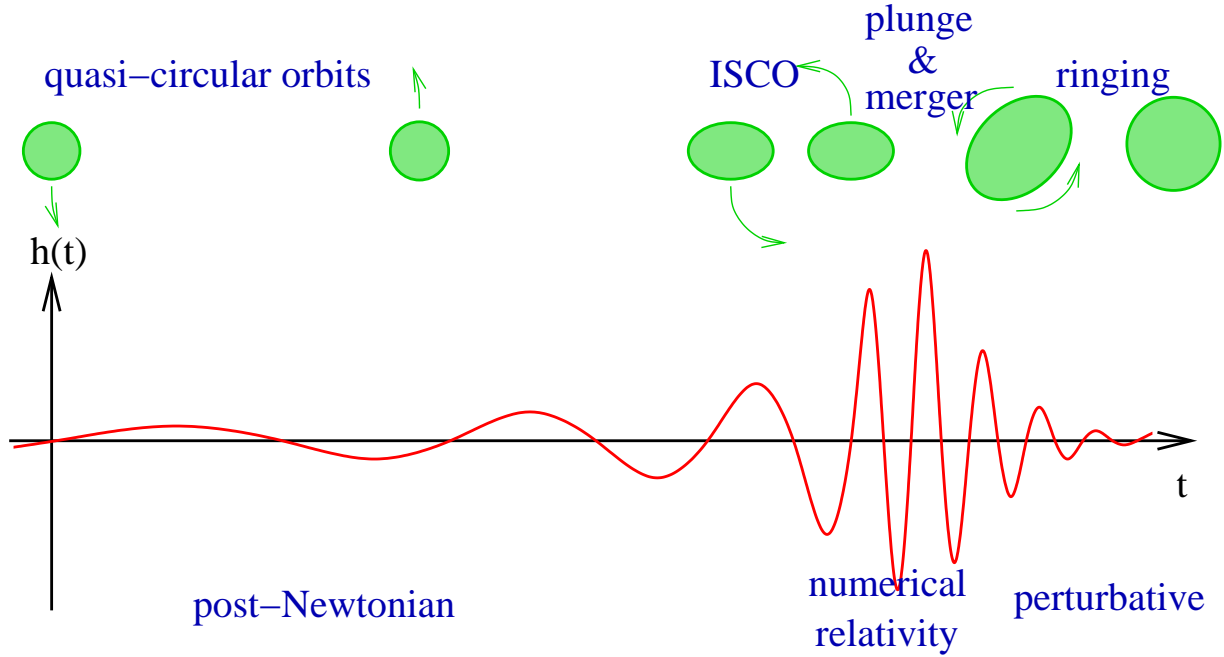}
\caption{Cartoon describing the inspiral of a black hole binary.}
\label{Fig4}
\end{figure}

To estimate $L_{\rm GW}$ for a binary system we evaluate $\Ibar_{jk}$
for a Newtonian binary and find
\begin{equation} \label{L_GW}
L_{\rm GW} = \frac{32}{5} \,\frac{G^4}{c^5}\,\frac{M^3 \mu^2}{a^5}.
\end{equation}
Here $M = m_1 + m_2$ is the binary's total mass, $\mu = m_1m_2/M$ the
reduced mass and $a$ the semi-major axis, and we have also used
Kepler's law to eliminate the orbital frequency $\Omega$.  Inserting
numbers we find that the only binary systems that emit appreciable
amounts of gravitational radiation have huge masses and very small
binary separations $a$.  The most promising candidates are therefore
{\em compact binaries} consisting of black holes or neutron stars.

The loss of energy due to the emission of gravitational radiation
leads a shrinking of the binary's orbit.  To see this we compute
\begin{equation} \label{J_drdt}
\frac{da}{dt} = \frac{dE/dt}{dE/da} = - \frac{L_{\rm GW}}{dE_{\rm bind}/da}
= - \frac{64}{5} \left( \frac{G}{c^2} \, 
\frac{M^{2/3}\mu^{1/3}}{a} \right)^3 c,
\end{equation}
where
\begin{equation}
E_{\rm bind} = \frac{1}{2} \,\frac{G M \mu}{a}
\end{equation}
is the generalization of (\ref{Newt_bind}).  Computing the loss of
angular momentum we would also find that that the emission of
gravitational radiation leads to a reduction of the binary's eccentricity,
i.e.~to a circularization of its orbit.

A binary inspiral then proceeds as illustrated in Fig.~\ref{Fig4}.
Presumably the binary starts out at a large binary separation.  The
emission of gravitational radiation leads to a continuous decrease in
the binary separation, and also to a decrease in the binary's
eccentricity.  At sufficiently late times we may therefore approximate
the binary orbit as circular, except for the slow inspiral.  As the
binary separation shrinks, both the amplitude and the frequency of the
emitted gravitational wave signal increases.  This leads to the
typical ``chirp'' signal sketched in Fig.~\ref{Fig4}.  In analogy to
our discussion of point masses orbiting a single black hole, the
binary will at some point reach an ISCO.  Inside this separation it is
energetically favorable for the binary companions to abandon circular
orbits, and instead plunge toward each other and merge.  In the final
``ring-down'' phase the remnant will settle down quickly into an
axisymmetric equilibrium object.

Even for the most promising sources of gravitational radiation any
astrophysical signal that we might hope for is going to be extremely
weak.  To see this we estimate the effect of a gravitational wave on
the spacetime metric $g_{ab}$.  If the spacetime is almost flat, we
may write $g_{ab} = \eta_{ab} + h_{ab}$, where $\eta_{ab}$ is the flat
Minkowski metric and where $h_{ab}$ is the small perturbation that we
observe as a gravitational wave.  For a perturbation caused by a
distant gravitational wave source we have
\begin{equation}
h_{ij}^{\rm TT} = \frac{2}{s} \, \frac{G}{c^4} \, \ddot {\Ibar}_{ij}^{\rm TT}
\left( t - \frac{s}{c} \right).
\end{equation}
Here the symbol ${\rm TT}$ denotes the ``transverse traceless'' part
of the corresponding tensors, $s$ is the distance from the observer to
the gravitational wave source, and the quadrupole moment
${\Ibar}_{ij}$ is evaluated at the retarded time $t - s/c$.  We again
evaluate this term for a Newtonian binary, and estimate the
gravitational wave amplitude to be in the order of
\begin{equation} \label{h}
h \approx \frac{1}{s} \, \frac{G}{c^4} \, M a^2 \Omega^2
\approx  \frac{G^2}{c^4} \, \frac{M}{s} \, \frac{M}{a} \approx 
\frac{r_{\rm S}}{s} \, \frac{r_{\rm S}}{a},
\end{equation}
where we have used Kepler's third law to eliminate $\Omega$ and where
$r_{\rm S}$ is the Schwarzschild radius (\ref{r_ss}) of the mass $M$.
Evidently we expect the strongest gravitational wave signal $h_{ij}$
when the binary's semi-major axis $a$ is small, i.e.~close to the ISCO
where $a \approx r_{\rm S}$ (if we neglect factors of a few).

Consider, for example, a stellar-mass compact binary that is
coalescing somewhere in the Virgo cluster.  For such a binary, $r_{\rm
S}$ is in the order of a few km, and the distance $s$ is about 15 Mpc
or a few $10^{20}$ km.  We then have $h \approx 10^{-20}$.  Since the
corrections $h_{ij}$ are added to the background metric, which is of
order unity, the relative change in the spacetime metric is
exceedingly small, and the proposal to measure these perturbations
remarkably ambitious.

The weakness of any gravitational wave signal that we may
realistically expect has two important consequences for the prospect
of observing them: we need extremely sensitive gravitational wave
detectors, and we need accurate models of any potential sources to aid
in the identification of any signal in the noisy output of the
detector.  We will discuss both of these aspects in the following two
Sections.

\subsection{Gravitational wave detectors}

\begin{figure}[t]
\includegraphics[width=5in]{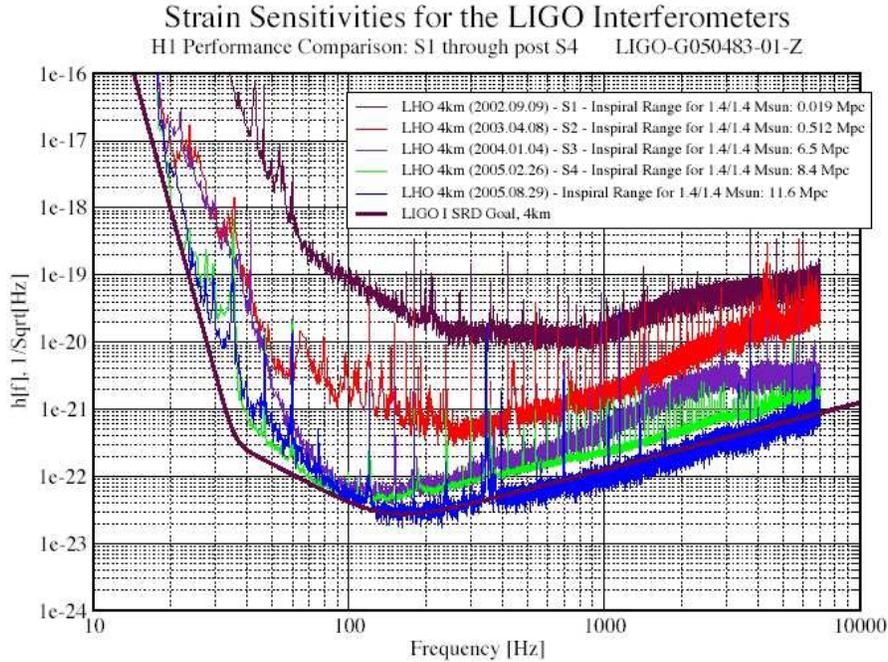}
\caption{Strain sensitivities for the LIGO Hanford (Washington)
interferometer (Figure courtesy the LIGO Scientific Collaboration).}
\label{Fig5}
\end{figure}

A new generation of gravitational wave observatories is now
operational.  Basically, these instruments are gigantic
Michelson-Morley interferometers.  The two LIGO observatories in the
US, for example, have an arm-length of 4 km.  Somewhat smaller
instruments exist in Italy (VIRGO), Germany (GEO), and Japan (TAMA),
and another observatory is currently under construction in Australia
(ACIGA).  The idea is that a gravitational wave $h_{ij}$ passing
through these interferometers will slightly distort the relative
length of the two perpendicular arms.  Tracking these distortions with
the help of laser interferences as a function of time should reveal
the passing gravitational wave signal.

The ground-based detectors mentioned above all have arm-lengths in the
order of a kilometer, which makes them sensitive to the gravitational
radiation emitted from stellar-mass black holes\footnote{In geometric
units a solar mass equals 1.48 km.}.  A space-based gravitational wave
antenna LISA with an arm-length of several million kilometer is being
planned; this instrument would be sensitive to gravitational radiation
emitted from supermassive black holes.

As we have seen in the previous section an astrophysical gravitational
wave signal that we might realistically expect will lead to a tiny
perturbation of the spacetime metric and hence to only a minuscule
distortion in the arm-lengths of the interferometers -- in fact, only
a tiny fraction of the size of the nucleus of a hydrogen atom.  It is
therefore extremely difficult to reduce the dominant sources of noise
-- seismic, thermal and photon shot-noise -- and increase the
signal-to-noise ratio so that an astrophysical source can be
identified unambiguously.  

It is therefore truly remarkable that the LIGO collaboration has
recently achieved its design goal of a strain sensitivity exceeding
$10^{-22}$ for a large part of its frequency range (see
Fig.~\ref{Fig5}).  Given this sensitivity, and assuming a
signal-to-noise ratio of eight, LIGO is now able to detect binary
neutron stars to a distance somewhat greater than 10 Mpc.  For binary
black holes with a slightly larger mass this range is also slightly
larger, and already includes the Virgo cluster.  

Whether or not the current LIGO observatory, LIGO I, will be able to
detect a binary black hole system -- or any other source of
gravitational radiation -- depends on how often such binaries coalesce
in our or our neighboring galaxies.  Estimates for binary neutron star
systems can be based on the statistics of known binaries\footnote{See,
e.g.~\cite{Kimetal05,Kaletal04}.}, while estimates for binary black
hole systems, which have never been observed, are typically based on
population synthesis calculations\footnote{For example
\cite{VosT03}.}.  According to these estimates it is not completely
impossible that LIGO I will observe a compact binary, and it seems
almost certain that an advanced LIGO II observatory, as it is
currently being planned, will see many such sources.

We finally point out that a gravitational wave detector tracks the
amplitude of a gravitational wave at one particular point in space,
but, unlike a telescope, does not produce an image.  A single detector
therefore cannot position any source.  Two detectors can position a
source to a ring in the sky, and with the world-wide network of
detectors there is hope that any potential source can be positioned to
within a reasonable accuracy.

\subsection{Source modeling}

Different approximations can be used to model the inspiral of a
compact binary in its different phases.  For the initial inspiral,
while the binary separation is sufficiently large and the effects of
relativistic and tidal interactions sufficiently small, post-Newtonian
point-mass calculations provide excellent approximations\footnote{See,
e.g., \cite{Bla02}, as well as L.~Blanchet's article in this volume.}.
The very late stage can be modeled very accurately with the help of
perturbation theory\footnote{E.g.~\cite{BakBCLT01}.}.  Neither one of
these approximations can be used in the intermediate regime around the
ISCO during which the binary emits the strongest gravitational wave
signal.  The most promising tool for the modeling of this dynamical
phase of the binary coalescence and merger is numerical
relativity\footnote{See \cite{BauS03} for a recent review.}.

Numerical relativity calculations typically adopt a 3+1 decomposition.
In such a decomposition the four-dimensional space $M$ is carved up
into a foliation of three-dimensional spatial slices $\Sigma$, each
one of which corresponds to an instant of constant coordinate time
$t$.  Einstein equations for the four-dimensional spacetime metric
$g_{ab}$ can then be rewritten as a set of three-dimensional equations
for the three-dimensional, spatial metric $\gamma_{ij}$ on the spatial
slices $\Sigma$, as well as its time derivative.

The general coordinate freedom of general relativity has an important
consequence for the structure of these three-dimensional equations.
Since there are three space coordinates and one time coordinate, we
can choose four of the ten independent components of the symmetric
spacetime metric $g_{ab}$ freely.  But that means that the ten
equations in Einstein's equations (\ref{field_gr}) cannot be
independent -- otherwise the metric would be over-determined.  Four of
the ten equations in Einstein's equations must be redundant.  In the
framework of a 3+1 decomposition these four equations are {\em
constraint equations} that constrain the gravitational fields within
each spatial slice $\Sigma$, while the remaining six {\em evolution
equations} govern the time-evolution of $\gamma_{ij}$ from one slice
$\Sigma$ to the next.  The equations are compatible, meaning that
fields that satisfy the constraint equations at one instant of time
continue to satisfy the constraints at all later times if the fields are
evolved with the evolution equations.  This structure of the equations
is very similar to that of Maxwell's equations, where the ``div''
equations constrain the electric and magnetic field at any instant of
time, while the ``curl'' equations govern the dynamical evolution of
the fields.

Finding a numerical solution to Einstein's equations usually proceeds in
two steps.  In the first step the constraint equations are solved to
construct {\em initial data}, describing the gravitational fields
together with any matter or other sources at some initial time, and in
the second step these data are {\em evolved} forward in time by
solving the evolution equations.

One of the challenges in constructing initial data results from the
fact that the constraint equations determine only some of the
gravitational fields.  The remaining fields are related to the degrees
of freedom associated with gravitational waves, which depend on the
past history and cannot be determined from Einstein's equations at only
one instant of time.  Instead, these ``background fields'' are freely
specifiable and have to be chosen before the constraint equations can
be solved.  The challenge, then, lies in making appropriate choices
that reflect the astrophysical scenario one wishes to model.  

For the modeling of binary black holes we would like to construct
initial data that model a binary at a reasonably close binary
separation outside the ISCO as it emerges from the inspiral from a
much larger separation.  We expect such a binary to be in
``quasi-equilibrium'', meaning that the orbit is circular (except for
the slow inspiral) and that the individual black holes are in
equilibrium.  A number of different approaches have been pursued, but
currently the best approximations to quasi-equilibrium black holes are
the models of \citet{CooP04}.  Probably there are ways to
improve these data, in particular as far as the background fields and
the resulting ``gravitational wave content'' are
concerned\footnote{One approach is to match the background fields to
the post-Newtonian approximation of the prior inspiral
\cite{TicBCD03}; another promising suggestion is a ``wave-less''
approximation proposed by \citet{ShiUF04}.}, but even without these
improvements the current models are probably excellent approximations.

\begin{figure}[t]
\includegraphics[width=3.1in]{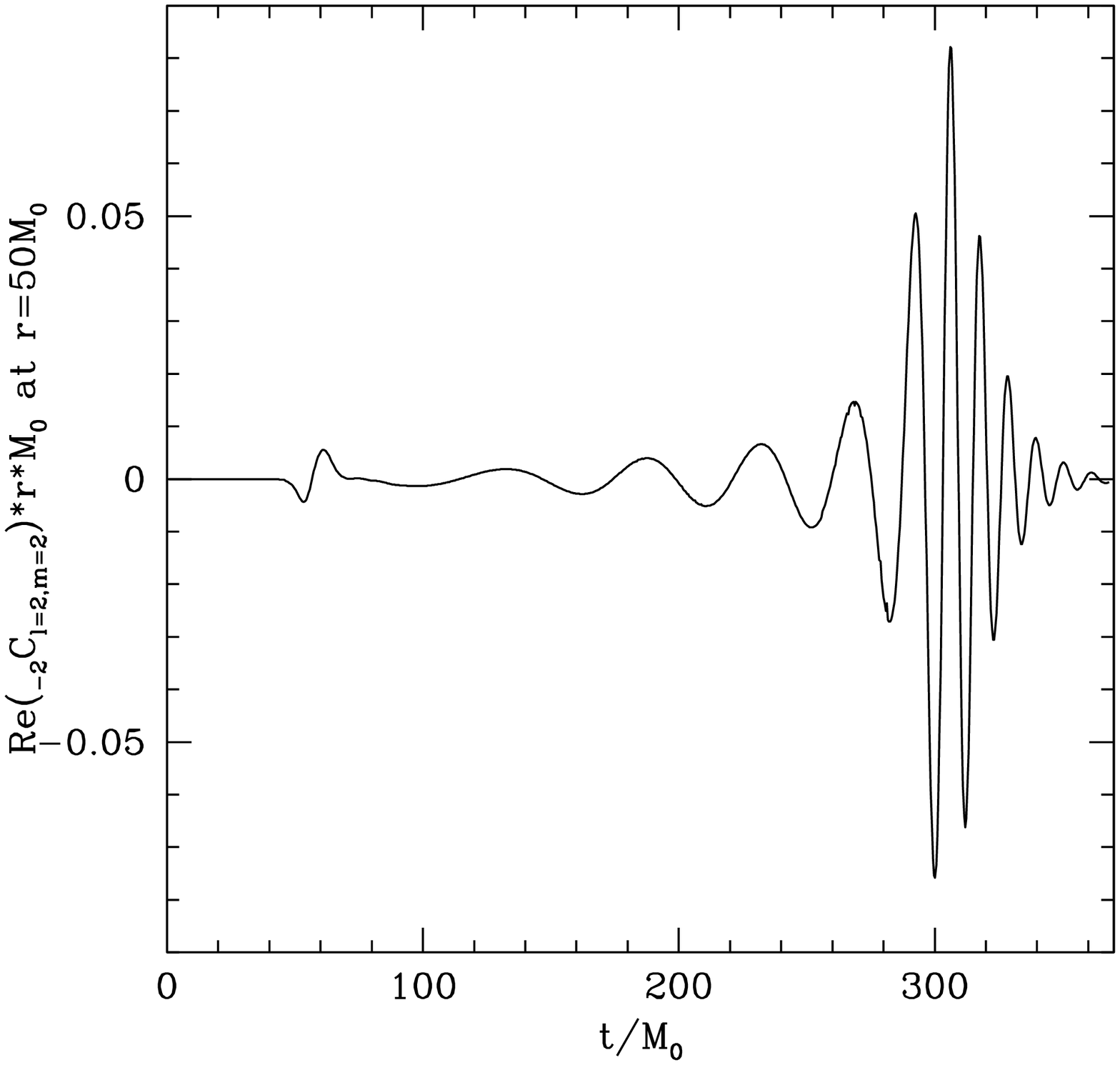}
\includegraphics[width=3.5in]{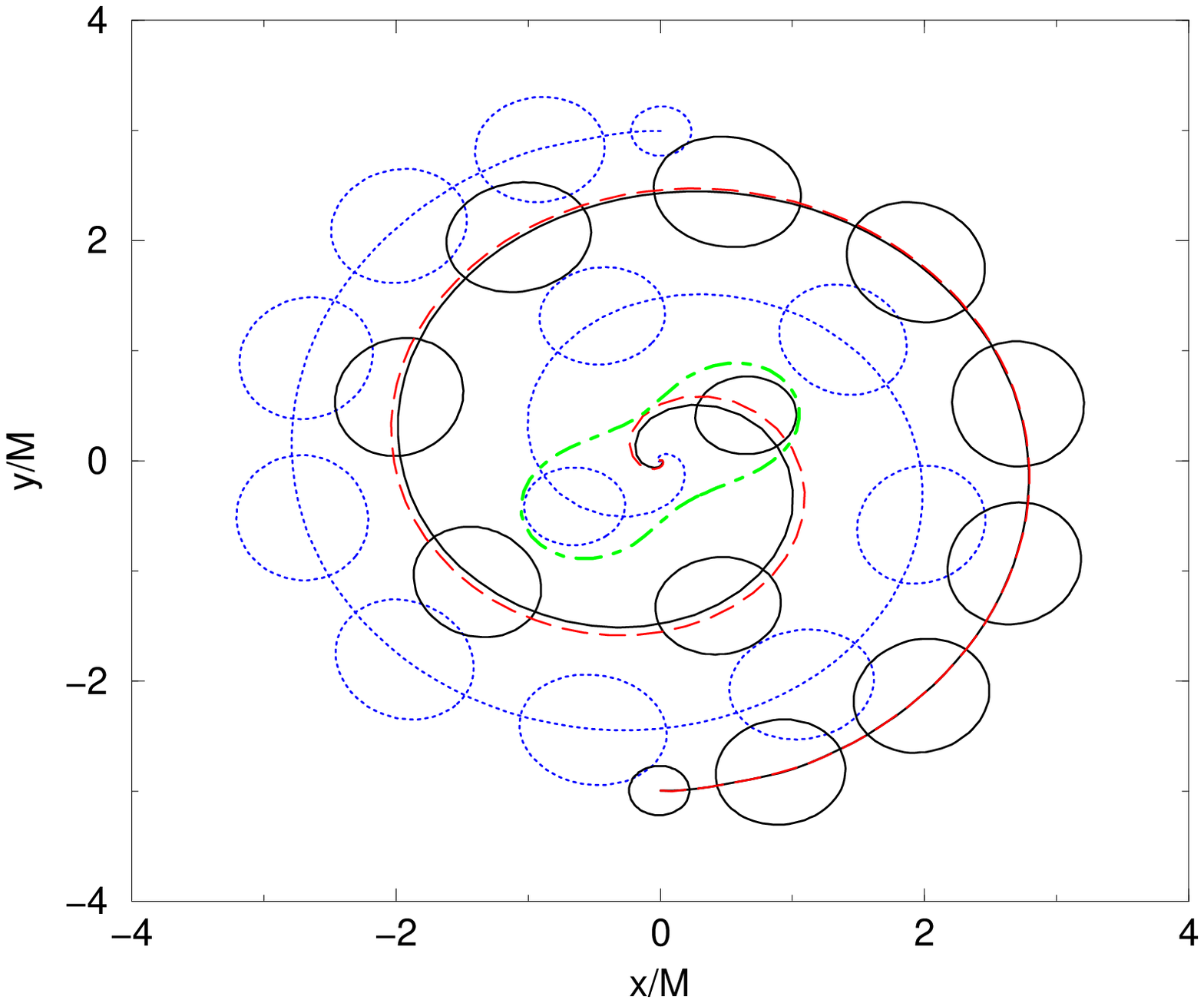}
\caption{The left panel shows the gravitational wave form Re($\psi_4$)
from a binary black hole coalescence as recently computed by Pretorius
(private communication).  The right panel shows the trajectory of the
black hole horizons in the calculation of \citet{CamLZ06}.}
\label{Fig6}
\end{figure}

While initial data describing binary black holes have been reasonably
well understood for a number of years, progress on dynamical
simulations of black holes has been much slower.  Until recently,
these simulations had been plagued by numerical instabilities that
made the codes crash long before the black holes had done anything
interesting.  The past year, however, has seen a dramatic
break-through in this field, and by now several groups can perform
reliable simulations of the binary black hole coalescence, merger and
ring-down.  

The first announcement of such a calculation came from \citet{Pre05}.
His approach differs from the more traditional numerical relativity
calculations in that he integrates the four-dimensional Einstein
equations (\ref{field_gr}) directly, instead of casting them into a
3+1 form\footnote{The equations are expressed in a way, however, that
leads to particularly desirable mathematical properties; see
\cite{Fri85,Gar02}.}.  The left panel of Fig.~\ref{Fig6} shows a
gravitational wave form from one of his recent simulations, starting
with the initial data of \cite{CooP04}.  At early times some noise is
seen, which is either related to the imperfect initial data themselves
or the imperfect matching of the numerical methods used in the initial
data and the evolution.  The noise disappears quickly, and leaves
behind a very clean waveform, tracking the inspiral through several
orbits, through the plunge and merger, until the newly formed remnant
settles down into a single Kerr black hole.

Shortly after Pretorius' announcement a number of other groups
announced similarly successful calculations
\cite{CamLMZ05,BakCDKvM05,DieHPSSTTV05}.  All of these later
simulations do adopt a 3+1 decomposition of Einstein's equations, and
cast these equations in the BSSN form \cite{ShiN95,BauS99}.  The right
panel of Fig.~\ref{Fig6} shows the trajectory of the black holes in
the calculation of \citet{CamLZ06}.  \citet{CamLMZ05,CamLZ06} and
\citet{BakCDKvM05} adopt a particularly simple method that treats the
black hole singularities as ``punctures'' and avoids having to excise
the black hole interior from the numerical grid \cite{BraB97,Bau00}.
In the meantime several follow-ups on these calculations have already
been published, including simulations of binaries with mass-ratios
different from unity \cite{HerSL06,BakCDKvMM06} and spinning black
holes \cite{CamLZ06b}.

Especially comparing with the situation just a year ago, it is truly
remarkable and reassuring that different groups using independent
techniques and implementations can now carry out reliable simulations
of binary black hole coalescence and merger. It may still be a while
until results from these simulations can be used to assemble a
catalog of realistic wave templates to be used in the analysis of
data from gravitational wave detectors, but the past year has
certainly seen a huge step forward in that direction.

\section{Summary}

Black holes may well be the most fascinating consequence of Einstein's
theory of general relativity.  Similarly fascinating is the
development of our understanding of black holes.  Speculations on
so-called ``dark stars'' actually predate both special and general
relativity by over a century, but it nevertheless took almost half a
century after the publication of general relativity and
Schwarzschild's derivation of his famous solution until black holes
became generally accepted as an astrophysical reality.

By now we have very convincing observational evidence for both
stellar-mass and supermassive black holes.  So far these observations
only constrain the black hole mass, and clearly it would be desirable
to probe the local black hole geometry in addition to the global
parameters.  There is hope that we will be able to do that in the near
future with the new generation of gravitational wave detectors.  The
LIGO observatories have recently achieved their design sensitivities,
enabling us to detect inspiraling binary black holes to a distance of
approximately the Virgo Cluster.  The next generation Advanced LIGO
will improve the sensitivity by over a factor of ten, which increases
the event rate by over a factor of thousand.  With the recent advances
in numerical relativity we are also much closer to producing
theoretical gravitational wave templates, which will aid both in the
identification of gravitational wave signals and in their
interpretation.  There is hope, then, that we can discuss detailed
gravitational wave observations of binary black holes by the time we
celebrate the centennial of Einstein's general relativity.

\begin{theacknowledgments}
It is a pleasure to thank Karen A.~Topp for a careful reading of this
manuscript.  This article was supported in part by NSF Grant
PHY-0456917 to Bowdoin College.
\end{theacknowledgments}




\begin{thebibliography}{60}
\expandafter\ifx\csname natexlab\endcsname\relax\def\natexlab#1{#1}\fi
\providecommand{\enquote}[1]{``#1''}
\expandafter\ifx\csname url\endcsname\relax
  \def\url#1{\texttt{#1}}\fi
\expandafter\ifx\csname urlprefix\endcsname\relax\def\urlprefix{URL }\fi
\providecommand{\eprint}[2][]{\url{#2}}

\bibitem[{Mitchell}(1796)]{Mit1796}
J.~{Mitchell}, \emph{Phil. Trans. R. Soc. Lond.} \textbf{141}, 1232 (1796).

\bibitem[{Laplace}(l'an IV de la R\'epublique Fran\c{c}aise (otherwise known as
  1795))]{Lap1795}
P.~S. {Laplace}, \emph{Exposition du systeme du monde}, L'imprimerie du
  Cercle-Social, Paris, l'an IV de la R\'epublique Fran\c{c}aise (otherwise
  known as 1795).

\bibitem[{Israel}(1973)]{Isr73}
W.~{Israel}, \enquote{Dark stars: the evolution of an idea,} in \emph{300 years
  of gravitation}, edited by S.~Hawking, and W.~Israel, Cambridge University
  Press, Cambridge, 1973.

\bibitem[{Einstein}(1915)]{Ein15}
A.~{Einstein}, \emph{Sitzungsberichte the Preussischen Akademie der
  Wissenschaften} pp. 844--847 (1915).

\bibitem[{Schwarzschild}(1916)]{Sch16}
K.~{Schwarzschild}, \emph{Sitzungsberichte the Deutschen Akademie der
  Wissenschaften} pp. 189--196 (1916).

\bibitem[{Oppenheimer} and {Snyder}(1939)]{OppS39}
K.~{Oppenheimer}, and H.~{Snyder}, \emph{Phys. Rev.} \textbf{56}, 455 (1939).

\bibitem[{Kruskal}(1960)]{Kru60}
M.~D. {Kruskal}, \emph{Phys. Rev.} \textbf{119}, 1743 (1960).

\bibitem[{Painlev\'e}(1921)]{Pai21}
P.~{Painlev\'e}, \emph{C. R. Acad. Sci. (Paris)} \textbf{173}, 677 (1921).

\bibitem[{Gullstrand}(1922)]{Gul22}
A.~{Gullstrand}, \emph{Arkiv Mat. Astron. Fys.} \textbf{16}, 1 (1922).

\bibitem[{Lema\^itre}(1933)]{Lem33}
G.~{Lema\^itre}, \emph{Ann. Soc. Sci. (Bruxelles) A} \textbf{53}, 51 (1933).

\bibitem[{Kerr}(1963)]{Ker63}
R.~P. {Kerr}, \emph{Phys. Rev. Lett.} \textbf{11}, 237 (1963).

\bibitem[{May} and {White}(1966)]{MayW66}
M.~W. {May}, and R.~H. {White}, \emph{Phys. Rev.} \textbf{74}, 35 (1966).

\bibitem[{Penrose}(1965)]{Pen65}
R.~{Penrose}, \emph{Phys. Rev. Lett.} \textbf{14}, 57 (1965).

\bibitem[{Israel}(1967)]{Isr67}
W.~{Israel}, \emph{Phys. Rev.} \textbf{164}, 1776 (1967).

\bibitem[{Carter}(1971)]{Car71}
B.~{Carter}, \emph{Phys. Rev. Lett.} \textbf{26}, 331 (1971).

\bibitem[{Robinson}(1975)]{Rob75}
D.~C. {Robinson}, \emph{Phys. Rev. Lett.} \textbf{34}, 905 (1975).

\bibitem[{Hazard} et~al.(1963)]{HazMS63}
C.~{Hazard}, M.~B. {Mackey}, and A.~J. {Shimmins}, \emph{Nature} \textbf{197},
  1037 (1963).

\bibitem[{Schmidt}(1963)]{Sch63}
M.~{Schmidt}, \emph{Nature} \textbf{197}, 1040 (1963).

\bibitem[{Oke}(1963)]{Oke63}
J.~B. {Oke}, \emph{Nature} \textbf{197}, 1040 (1963).

\bibitem[{Greenstein} and {Matthews}(1963)]{GreM63}
J.~{Greenstein}, and T.~A. {Matthews}, \emph{Nature} \textbf{197}, 1041 (1963).

\bibitem[{Hoyle} and {Fowler}(1963)]{HoyF63}
F.~{Hoyle}, and W.~A. {Fowler}, \emph{Nature} \textbf{197}, 533 (1963).

\bibitem[{Ginzburg}(1961)]{Gin61}
V.~L. {Ginzburg}, \emph{Sov. Astron.} \textbf{5}, 282 (1961).

\bibitem[{Oda}(1977)]{Oda77}
M.~{Oda}, \emph{Space Sci. Rev.} \textbf{20}, 757 (1977).

\bibitem[{Shapiro} and {Teukolsky}(1983)]{ShaT83}
S.~L. {Shapiro}, and S.~A. {Teukolsky}, \emph{Black Holes, White Dwarfs and
  Neutron Stars: The Physics of Compact Objects}, Wiley-Interscience, New York,
  1983.

\bibitem[{Gies} and {Bolton}(1982)]{GieB82}
D.~R. {Gies}, and C.~T. {Bolton}, \emph{Astrophys. J.} \textbf{260}, 240
  (1982).

\bibitem[{Baumgarte} et~al.(2000)]{BauSS00}
T.~W. {Baumgarte}, S.~L. {Shapiro}, and M.~{Shibata}, \emph{Astrophys. J.
  Lett.} \textbf{528}, L29 (2000).

\bibitem[{McClintock} and {Remillard}(2006)]{McCR06}
J.~E. {McClintock}, and R.~A. {Remillard}, \enquote{Black hole binaries,} in
  \emph{Compact Stellar X-ray Sources}, edited by W.~H.~G. {Lewin}, and M.~{van
  der Klis}, Cambridge University Press, Cambridge, 2006.

\bibitem[{Sch{\"o}del} et~al.(2002)]{Schetal02}
R.~{Sch{\"o}del}, T.~{Ott}, R.~{Genzel}, R.~{Hofmann}, M.~{Lehnert},
  A.~{Eckart}, N.~{Mouawad}, T.~{Alexander}, M.~J. {Reid}, R.~{Lenzen},
  M.~{Hartung}, F.~{Lacombe}, D.~{Rouan}, E.~{Gendron}, G.~{Rousset}, A.-M.
  {Lagrange}, W.~{Brandner}, N.~{Ageorges}, C.~{Lidman}, A.~F.~M. {Moorwood},
  J.~{Spyromilio}, N.~{Hubin}, and K.~M. {Menten}, \emph{Nature} \textbf{419},
  694 (2002).

\bibitem[{Ghez} et~al.(2003)]{Gheetal03}
A.~M. {Ghez}, G.~{Duch{\^e}ne}, K.~{Matthews}, S.~D. {Hornstein}, A.~{Tanner},
  J.~{Larkin}, M.~{Morris}, E.~E. {Becklin}, S.~{Salim}, T.~{Kremenek},
  D.~{Thompson}, B.~T. {Soifer}, G.~{Neugebauer}, and I.~{McLean},
  \emph{Astrophys. J. Lett.} \textbf{586}, L127 (2003).

\bibitem[{Miyoshi} et~al.(1995)]{Miyetal95}
M.~{Miyoshi}, J.~{Moran}, J.~{Herrnstein}, L.~{Greenhill}, N.~{Nakai},
  P.~{Diamond}, and M.~{Inoue}, \emph{Nature} \textbf{373}, 127 (1995).

\bibitem[{Richstone} et~al.(1998)]{Ricetal98}
D.~{Richstone}, E.~A. {Ajhar}, R.~{Bender}, G.~{Bower}, A.~{Dressler}, S.~M.
  {Faber}, A.~V. {Filippenko}, K.~{Gebhardt}, R.~{Green}, L.~C. {Ho},
  J.~{Kormendy}, T.~R. {Lauer}, J.~{Magorrian}, and S.~{Tremaine},
  \emph{Nature} \textbf{395}, A14 (1998).

\bibitem[{Rees}(1984)]{Ree84}
M.~{Rees}, \emph{Ann. Rev. Astr. \& Ap.} \textbf{22}, 471 (1984).

\bibitem[{Fan} et~al.(2003)]{Fanetal03}
X.~{Fan}, M.~A. {Strauss}, D.~P. {Schneider}, R.~H. {Becker}, R.~L. {White},
  Z.~{Haiman}, M.~{Gregg}, L.~{Pentericci}, E.~K. {Grebel}, V.~K. {Narayanan},
  Y.-S. {Loh}, G.~T. {Richards}, J.~E. {Gunn}, R.~H. {Lupton}, G.~R. {Knapp},
  {\v Z}.~{Ivezi{\'c}}, W.~N. {Brandt}, M.~{Collinge}, L.~{Hao}, D.~{Harbeck},
  F.~{Prada}, J.~{Schaye}, I.~{Strateva}, N.~{Zakamska}, S.~{Anderson},
  J.~{Brinkmann}, N.~A. {Bahcall}, D.~Q. {Lamb}, S.~{Okamura}, A.~{Szalay}, and
  D.~G. {York}, \emph{Astron. J.} \textbf{125}, 1649 (2003).

\bibitem[{Shapiro}(2005)]{Sha05}
S.~L. {Shapiro}, \emph{Astrophys. J.} \textbf{620}, 59 (2005).

\bibitem[{Volonteri} and {Rees}(2005)]{VolR05}
M.~{Volonteri}, and M.~{Rees}, \emph{Astrophys. J.} \textbf{633}, 624 (2005).

\bibitem[{Najayan}(2005)]{Nar05}
R.~{Najayan}, \emph{New J. Phys.} \textbf{7}, 199 (2005).

\bibitem[{Miller} et~al.(2004)]{Miletal04}
J.~M. {Miller}, A.~C. {Fabian}, C.~S. {Reynolds}, M.~A. {Nowak}, J.~{Homan},
  M.~J. {Freyberg}, M.~{Ehle}, T.~{Belloni}, R.~{Wijnands}, M.~{van der Klis},
  P.~A. {Charles}, and W.~H.~G. {Lewin}, \emph{Astrophys. J. Lett.}
  \textbf{606}, L131 (2004).

\bibitem[{Kim} et~al.(2005)]{Kimetal05}
C.~{Kim}, V.~{Kalogera}, D.~R. {Lorimer}, M.~{Ihm}, and K.~{Belczynski},
  \enquote{{The Galactic Double-Neutron-Star Merger Rate: Most Current
  Estimates},} in \emph{ASP Conf. Ser. 328: Binary Radio Pulsars}, edited by
  F.~A. {Rasio}, and I.~H. {Stairs}, 2005, p.~83.

\bibitem[{Kalogera} et~al.(2004)]{Kaletal04}
V.~{Kalogera}, C.~{Kim}, D.~R. {Lorimer}, M.~{Burgay}, N.~{D'Amico},
  A.~{Possenti}, R.~N. {Manchester}, A.~G. {Lyne}, B.~C. {Joshi}, M.~A.
  {McLaughlin}, M.~{Kramer}, J.~M. {Sarkissian}, and F.~{Camilo},
  \emph{Astrophys. J. Lett.} \textbf{601}, L179 (2004).

\bibitem[{Voss} and {Tauris}(2003)]{VosT03}
R.~{Voss}, and T.~M. {Tauris}, \emph{Mon. Not. Roy Astron. Soc.} \textbf{342},
  1169 (2003).

\bibitem[{Blanchet}(2002)]{Bla02}
L.~{Blanchet}, \emph{Living Rev. Relativity} \textbf{5}, 3 (2002).

\bibitem[{Baker} et~al.(2001)]{BakBCLT01}
J.~{Baker}, B.~{Br{\"u}gmann}, M.~{Campanelli}, C.~O. {Lousto}, and
  R.~{Takahashi}, \emph{Phys. Rev. Lett.} \textbf{87}, 121103 (2001).

\bibitem[{Baumgarte} and {Shapiro}(2003)]{BauS03}
T.~W. {Baumgarte}, and S.~L. {Shapiro}, \emph{Phys. Rept.} \textbf{376}, 41
  (2003).

\bibitem[{Cook} and {Pfeiffer}(2004)]{CooP04}
G.~B. {Cook}, and H.~P. {Pfeiffer}, \emph{Phys. Rev. D} \textbf{70}, 104016
  (2004).

\bibitem[{Tichy} et~al.(2003)]{TicBCD03}
W.~{Tichy}, B.~{Br{\"u}gmann}, M.~{Campanelli}, and P.~{Diener}, \emph{Phys.
  Rev. D} \textbf{67}, 064008 (2003).

\bibitem[{Shibata} et~al.(2004)]{ShiUF04}
M.~{Shibata}, K.~{Ury{\= u}}, and J.~L. {Friedman}, \emph{Phys. Rev. D}
  \textbf{70}, 044044 (2004).

\bibitem[{Campanelli} et~al.(2006{\natexlab{a}})]{CamLZ06}
M.~{Campanelli}, C.~O. {Lousto}, and Y.~{Zlochower}, \emph{Phys. Rev. D}
  \textbf{73}, 061501 (2006{\natexlab{a}}).

\bibitem[{Pretorius}(2005)]{Pre05}
F.~{Pretorius}, \emph{Phys. Rev. Lett.} \textbf{95}, 121101 (2005).

\bibitem[{Friedrich}(1985)]{Fri85}
H.~{Friedrich}, \emph{Commun. Math. Phys.} \textbf{100}, 525 (1985).

\bibitem[{Garfinkle}(2002)]{Gar02}
D.~{Garfinkle}, \emph{Phys. Rev. D} \textbf{65}, 044029 (2002).

\bibitem[{Campanelli} et~al.(2005)]{CamLMZ05}
M.~{Campanelli}, C.~O. {Lousto}, P.~{Marronetti}, and Y.~{Zlochower}  (2005),
  \eprint{gr-qc/0511048}.

\bibitem[{Baker} et~al.(2005)]{BakCDKvM05}
J.~G. {Baker}, J.~{Centrella}, C.~{Dai-Il}, M.~{Koppitz}, and J.~R. {van Meter}
   (2005), \eprint{gr-qc/0511103}.

\bibitem[{Diener} et~al.(2005)]{DieHPSSTTV05}
P.~{Diener}, F.~{Herrmann}, D.~{Pollney}, E.~{Schnetter}, E.~{Seidel},
  R.~{Takahashi}, J.~{Thornburg}, and J.~{Ventrella}  (2005),
  \eprint{gr-qc/0512108}.

\bibitem[{Shibata} and {Nakamura}(1995)]{ShiN95}
M.~{Shibata}, and T.~{Nakamura}, \emph{Phys. Rev. D} \textbf{52}, 5428 (1995).

\bibitem[{Baumgarte} and {Shapiro}(1999)]{BauS99}
T.~W. {Baumgarte}, and S.~L. {Shapiro}, \emph{Phys. Rev. D} \textbf{59}, 024007
  (1999).

\bibitem[{Brandt} and {Br\"ugmann}(1997)]{BraB97}
S.~{Brandt}, and B.~{Br\"ugmann}, \emph{Phys. Rev. Lett.} \textbf{78}, 3606
  (1997).

\bibitem[{Baumgarte}(2000)]{Bau00}
T.~W. {Baumgarte}, \emph{Phys. Rev. D} \textbf{62}, 024018 (2000).

\bibitem[{Herrmann} et~al.(2006)]{HerSL06}
F.~{Herrmann}, D.~{Shoemaker}, and P.~{Laguna}  (2006), \eprint{gr-qc/0601026}.

\bibitem[{Baker} et~al.(2006)]{BakCDKvMM06}
J.~G. {Baker}, J.~{Centrella}, C.~{Dai-Il}, M.~{Koppitz}, J.~R. {van Meter},
  and M.~C. {Miller}  (2006), \eprint{astro-ph/0603204}.

\bibitem[{Campanelli} et~al.(2006{\natexlab{b}})]{CamLZ06b}
M.~{Campanelli}, C.~O. {Lousto}, and Y.~{Zlochower}  (2006{\natexlab{b}}),
  \eprint{gr-qc/0604012}.

\end{thebibliography}

\end{document}